# Primary Investigation on Ram-Rotor Detonation Engine

Haocheng Wen*, Bing Wang†
School of Aerospace Engineering, Tsinghua University, Beijing, China, 100084
* haochengwenson@126.com,  †wbing@tsinghua.edu.cn

**Abstract**

The study presents a new type of detonation engine called the Ram-Rotor Detonation Engine (RRDE), which overcomes some of the drawbacks of conventional detonation engines such as pulsed detonation engines, oblique detonation engines, and rotating detonation engines. The RRDE organizes the processes of reactant compression, detonation combustion, and burned gas expansion in a single rotor, allowing it to achieve an ideal detonation cycle under a wide range of inlet Mach numbers, thus significantly improving the total pressure gain of the propulsion system. The feasibility and performance of RRDE are discussed through theoretical analysis and numerical simulations. The theoretical analysis indicates that the performance of the RRDE is mainly related to the inlet velocity, the rotor rim velocity and the equivalence ratio of reactant. Increasing the inlet velocity leads to a decrease in the total pressure gain of the RRDE. Once the inlet velocity exceeds the critical value, the engine cannot achieve positive total pressure gain. Increasing the rim velocity can improve the total pressure gain and the thermodynamic cycle efficiency of RRDE. Increasing the equivalence ratio can also improve the thermodynamic cycle efficiency and enhance the total pressure gain at lower inlet velocities. While, at higher inlet velocities, increasing the equivalence ratio may reduce the total pressure gain. Numerical simulations are also performed to analyze the detailed flow field structure in RRDE and its variations with the inlet parameters. The simulation results demonstrate that the detonation wave can stably stand in the RRDE and can adapt to the change of the inlet equivalence ratio within a certain range. This study provides preliminary theoretical basis and design reference for the RRDE.

**Nomenclature**

- $p$     Static pressure
- $p_t$     Total pressure measured with respect to the body-fixed reference frame of RRDE
- $\rho$     Density
- $T$     Static temperature
- $c$     Sound speed
- $\gamma$     Specific heat ratio
- $R$     Specific gas constant
- $V$     Velocity scalar
- $Ma$     Mach number
- $\pi_c$     Compression ratio
- $\pi_e$     Expansion ratio
- $\theta_D$     Angle between detonation wave front and $x$-axis
- $\alpha$     Ratio of relative Mach number of inflow to Chapman-Jouguet Mach number of detonation
- $\beta$     Ratio of absolute Mach number of inflow to Chapman-Jouguet Mach number of detonation
- $\varphi$     Equivalence ratio

PG    Total pressure gain of the engine measured with respect to the body-fixed reference frame
Π    Total pressure ratio measured with respect to the body-fixed reference frame

*Subscript*

0    State at the inlet
1    State after the detonation wave
2    State at the outlet
D    State in front of the detonation wave
CJ    Chapman-Jouguet state

*Abbreviation*

RRDE    Ram-rotor detonation engine
DW    Detonation wave
SW    Shock wave
OSW    Oblique shock wave

## 1. Introduction

Detonation is a combustion wave propagating at supersonic speeds that consists of a precursor shock and an accompanying reaction zone. Typically, stable detonation waves can reach a Mach number of 4 to 6 relative to the incoming flow. Under the compression effect of the precursor shock, the static pressure ratio between the reaction zone and the wave front can reach 13 to 55 times [1], indicating the detonation is a typical form of pressure gain combustion. The idea of using detonation to improve thermodynamic cycle efficiency and enhance the performance of propulsion systems has been around for a long time. In 1940, Zel'dovich first theoretically analyzed the thermodynamic cycle efficiency and the performance of air-breathing detonation engines [2]. In 1957, Nicholls et al. [3] first proposed the concept of the pulsed detonation engine (PDE) for detonation propulsion systems and conducted experimental verification. Subsequently, they proposed the concepts of the standing detonation wave (SDW) engine [4] (which is later developed into the oblique detonation engine, ODE [5,6]) and the rotating detonation engine (RDE) [7], which are present typical types of detonation propulsion.

In the following decades, these three types of detonation propulsion gradually gained attention worldwide, and extensive related research has been conducted. The research on PDE started relatively early, but due to difficulties in achieving stable thrust at low operating frequencies, PDE has yet to reach the practical stage [8–10]. ODE requires high Mach numbers in the incoming flow, and the required experimental conditions are harsh, so numerical simulation studies are still predominant at current stage, with limited its research progress [11–13]. RDE, on the other hand, has become a research hotspot in the field of aerospace propulsion in recent years due to its advantages of simple structure, continuous thrust, and suitability for a wide range of flight Mach numbers [14–19]. Its attention is still continuously increasing, and a considerable amount of engineering exploration on RDE has already been conducted, including flight tests and space experiments [20,21].

However, although the detonation is a form of pressure gain combustion, it does not necessarily mean that applying detonation to a propulsion system will increase its total pressure gain or thermodynamic cycle efficiency. The theoretical analysis from Zel'dovich's early study [2] explicitly stated that applying detonation combustion in energy production does not provide significant gains compared to constant volume combustion. Additionally, in supersonic air-breathing jet engines with continuous combustion, even without considering other non-ideal losses, adopting detonation still leads to a decrease in thrust compared to traditional thermodynamic cycle modes [2]. Wintenberger et al. [22,23] further demonstrated that it is impossible for a standing detonation propulsion system to achieve a positive total pressure gain based on stagnation Hugoniot analysis. Positive total pressure gain can only be achieved with the application of propagating detonation waves [22,24].

As the currently most prominent type of detonation propulsion, RDE was initially believed to significantly enhance the total pressure gain of propulsion systems. As a form of propagating detonation waves, the rotating detonation wave has the potential to achieve the total pressure gain from the theoretical aspect. Extensive theoretical analyses and numerical simulation studies also suggest that rotating detonation engines can achieve positive total pressure gain [25,26]. However, there is currently few experimental evidence indicating the attainment of evident positive total pressure gain in rotating detonation combustors or engines [27–29].

To explain the contradiction between theoretical and experimental results, we proposed a quasi-one-dimensional theoretical model to discuss the relationship between the total pressure gain and systematic parameters in detonation propulsion, highlighting the necessity of employing specific configurations to achieve a positive total pressure gain in RDEs [30]. Furthermore, based on a comprehensive calculation model for the total pressure of RDEs, we elucidated the significant impact of reverse oblique shock waves induced by rotating detonation waves on reducing the total pressure gain [31]. Finally, we proposed a technical concept that utilizes ideal check-valves, compressors, and turbines to eliminate reverse oblique shock waves, thereby achieving the ideal total pressure gain [31].

Continuing this technical concept and drawing inspiration from the ram-rotor compressor, we propose a new conceptive Ram-Rotor Detonation Engine (RRDE). The ram-rotor compressor was initially proposed by Ramgen company [32], and it can accelerate the incoming flow to relative supersonic speeds through high-speed rotating rotors, followed by deceleration and compression of the flow in the compressor. The feasibility of the ram-rotor compressor has been extensively validated through numerical and experimental research [33–35]. In the RRDE, the compression, detonation combustion, and expansion of the reactant flow are successively organized between the blades of rotors. RRDE utilizes propagating detonation waves to achieving the ideal detonation thermodynamic cycle, thereby enhancing the propulsion performance.

This paper presents preliminary research on the RRDE. Firstly, the basic structure and principle are introduced, followed by a theoretical analysis of RRDEs performance. Finally, the feasibility of RRDE is validated through numerical simulations.

## 2. Conceptual design

The basic concept of the Ram-Rotor Detonation Engine (RRDE) is to achieve the continuous propagation of detonation waves and the sustained operation of the combustor, while minimizing the impact of non-isentropic shock wave during the compression and expansion processes. This enables the engine to approach the total pressure gain and thermodynamic efficiency of the ideal detonation cycle.

Figure 1 shows the typical structures of RRDEs. It mainly consists of a rotating rotor with blades, and a stationary casing. The rotor can input or output torque through the shafts. The blades on the rotor are distributed in a helical symmetric manner. The combustible mixture undergoes compression, detonation combustion, and expansion within the variable cross-sectional channels between the blades. There are various ways to implement the variable cross-sectional channels, including by varying the blade thickness as shown in Figure 1a, or by varying the radial dimensions of the rotor as shown in Figure 2b, or a combination of the above methods. As a preliminary study and without loss of generality, the subsequent analysis and discussion mainly focus on the RRDE type shown in Figure 1a.

To provide a clearer description of the flow process between the blades, the mid-height cross-section of the flow channel of RRDE shown in Figure 1a is extracted and unfolded along the circumferential direction, as shown in Figure 2. The rim velocity of the rotor at this cross-section is $V_{rot}$. The inlet velocity of the premixed combustible mixture is $V_{in}$. With respect to the moving reference frame fixed at the blades, the combustible mixture enters the flow channel with a relative velocity of $V_0$. After undergoing the compression process, the relative velocity of the mixture decreases to $V_D$ in front of the detonation wave, and thus the detonation wave keeps stationary relative to the blades. The high-temperature burned gas after the detonation wave undergoes expansion and exhausts at the outlet.

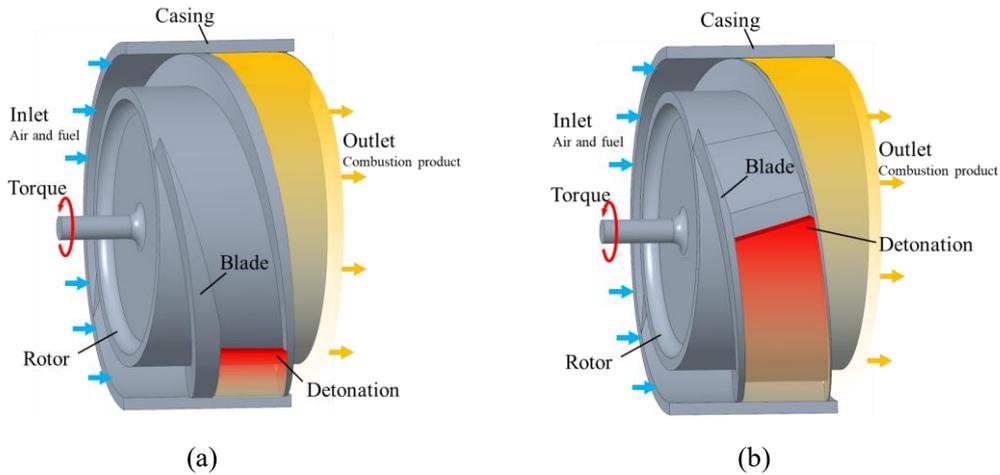

**Figure 1.** Typical structures of RRDEs. (a) scheme by varying the blade thickness; (b) scheme by varying the radial dimensions of the rotor.

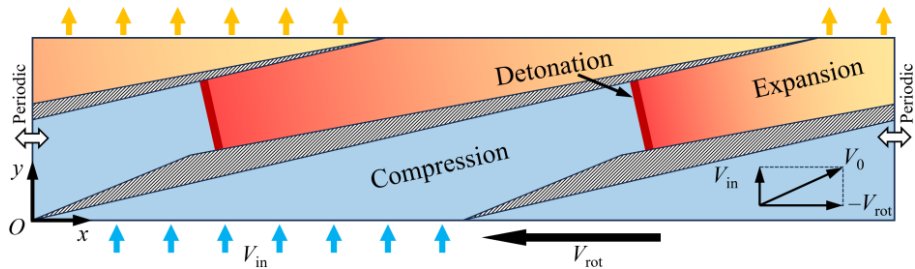

**Figure 2.** Mid-height cross-section of the RRDE flow channel which is unfolded along the circumferential direction.

The RRDE has the following significant advantages:
- The compression, combustion, and expansion processes of the combustible mixture are all accomplished in a single component. The detonation wave is employed to significantly reduce the distance of complete combustion, and thus resulting in a simple and compact system structure.

- The thermodynamic cycle efficiency and total pressure gain of the engine are significantly improved based on the continuously propagating detonation waves.
- By adjusting the rotor speed, RRDE can be applicable to a wide range of inlet Mach numbers.

### 3. Theoretical analysis

In this section, the basic performance of RRDE and its influence parameters are analyzed theoretically. Since the flow parameters vary in the radial direction, the parameters at the mid-height cross-section shown in Figure 3 are selected as the characteristic parameters for analysis. Additionally, the following assumptions are adopted: the flow is quasi-one-dimensional and inviscid; the detonation wave satisfies the Chapman-Jouguet (CJ) relation; the compression and expansion processes are isentropic, and the walls are adiabatic.

Firstly, the flow field parameters at each characteristic station are calculated, and then the propulsion performance and thermodynamic cycle efficiency of RRDE are analyzed.

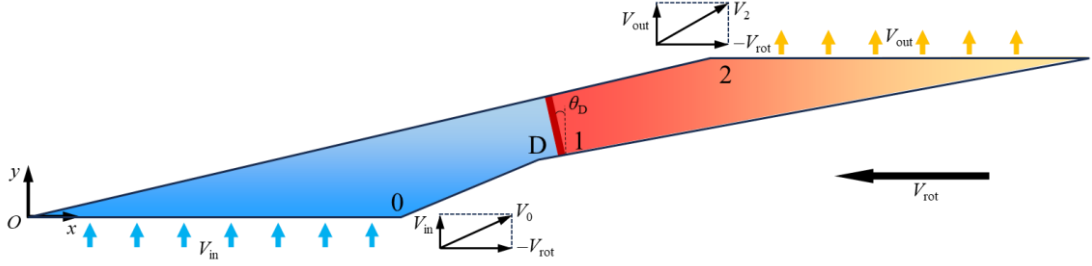

**Figure 3.** Characteristic stations and parameters defined on the mid-height cross-section of RRDE.

### 3.1 Flow Parameters in RRDE

As shown in Figure 3, we assume the velocities of inlet and outlet flows of RRDE have only axial components (i.e. $y$ direction) with respect to the body-fixed reference frame of RRDE, which are $V_{in}$ and $V_{out}$ respectively. It should be noted that under this assumption, the overall torque of the rotor is zero. Define subscripts 0, D, 1, and 2 to respectively represent the inlet station, the station in front of detonation wave, the station of CJ plane after detonation wave, and the outlet station. Select the moving reference frame fixed to the blade with the rim velocity of $V_{rot}$. Then the relative velocity is defined in the moving reference frame, and the absolute velocity is defined with respect to the body-fixed reference frame of RRDE.

#### 3.1.1 Parameters at station 0

Assume that the static pressure of inlet flow is $p_0$, the static temperature is $T_0$, the specific heat ratio is $\gamma_0$, the gas constant is $R_0$, the density is $\rho_0 = p_0/R_0 T_0$, and the sound speed is $c_0 = (\gamma_0 p_0/\rho_0)^{1/2}$. The absolute and relative Mach number of inlet flow are $Ma_{in}$ and $Ma_0$, and can be calculated by

$$Ma_{in} = \frac{V_{in}}{c_0}, \quad Ma_0 = \frac{\sqrt{V_{in}^2 + V_{rot}^2}}{c_0} \tag{1}$$

#### 3.1.2 Parameters at station D

In order to keep the detonation wave stationary relative to the blade, the Mach number in front of the detonation wave should be equal to the Mach number of CJ detonation, namely $Ma_D$. Assuming that the specific heat ratio is constant during the compression, that is $\gamma_D = \gamma_0$, the compression ratio $\pi_c$ of the flow in the compression stage is satisfied

$$\pi_c = \left( \frac{1 + \frac{\gamma_0 - 1}{2} Ma_0^2}{1 + \frac{\gamma_0 - 1}{2} Ma_D^2} \right)^{\frac{\gamma_0}{\gamma_0 - 1}} \tag{2}$$

Then the static pressure $p_D$, the density $\rho_D$ and the temperature $T_D$ in front of detonation wave are respectively

$$p_D = \pi_c p_0, \quad \rho_D = \pi_c^{\frac{1}{\gamma_0}} \rho_0, \quad T_D = \pi_c^{\frac{\gamma_0 - 1}{\gamma_0}} T_0 \tag{3}$$

According to Chapman-Jouguet theory, the detonation wave Mach number $Ma_D$ can be approximately calculated by [36]

$$Ma_D \approx \frac{1}{c_D} \sqrt{2(\gamma_1^2 - 1)q} \tag{4}$$

where $q$ is the difference between the enthalpies of formation of reactants and products, and $\gamma_1$ is the specific heat ratio of detonation products. $c_D$ is the sound speed of the reactant flow in front of the detonation wave, and is equal to

$$c_D = \sqrt{\gamma_0 R_0 T_0 \pi_c^{\frac{\gamma_0 - 1}{\gamma_0}}} \tag{5}$$

Combining Equations (2)-(5), $\pi_c$ can be expressed as

$$\pi_c = \left( 1 + \frac{\gamma_0 - 1}{2} Ma_0^2 - \frac{(\gamma_0 - 1)(\gamma_1^2 - 1)q}{\gamma_0 R_0 T_0} \right)^{\frac{\gamma_0}{\gamma_0 - 1}} \tag{6}$$

Define the angle between the detonation wave and the $y$-axis is $\theta_D$, then the flow in the compression stage satisfies the $x$-direction momentum conservation in the moving reference frame, and the energy conservation in the body-fixed reference frame, as follows

$$\dot{m} V_{rot} + F_x = \dot{m} V_D \cos \theta_D \tag{7}$$

$$\begin{aligned} F_x V_{rot} &= \dot{m} C_V T_{t0} \left( \frac{T_{tD}}{T_{t0}} - 1 \right) \\ &= \dot{m} \frac{c_0^2}{\gamma_0 (\gamma_0 - 1)} \left( \left( 1 + \frac{\gamma_0 - 1}{2} Ma_D'^2 \right) \pi_c^{\frac{\gamma_0 - 1}{\gamma_0}} - \left( 1 + \frac{\gamma_0 - 1}{2} Ma_{in}^2 \right) \right) \end{aligned} \tag{8}$$

where $F_x$ is the $x$-component of the force acting on the blade in the compression stage. $T_{t0}$ and $T_{tD}$ are the total temperature at station 0 and D with respect to the body-fixed reference frame. $C_V$ is the specific heat capacity at constant volume of the flow in the compression stage. $Ma'_D$ is the absolute Mach number of the flow in front of the detonation wave, and is equal to

$$Ma'_D = Ma_D \sqrt{\left( \cos \theta_D - \frac{V_{rot}}{Ma_D c_D} \right)^2 + \sin^2 \theta_D} \tag{9}$$

Combining Equations (7)-(9), $\theta_D$ can be calculated by

$$\theta_{\mathrm{D}} = \arccos \frac{\frac{c_0^2}{\gamma_0(\gamma_0-1)}\left(\left(1+\frac{\gamma_0-1}{2}Ma_{\mathrm{D}}^2\left(1+\left(\frac{V_{\mathrm{rot}}}{V_{\mathrm{D}}}\right)^2\right)\right)\pi_{\mathrm{c}}^{\frac{\gamma_0-1}{\gamma_0}} - \left(1+\frac{\gamma_0-1}{2}Ma_{\mathrm{in}}^2\right)\right)+V_{\mathrm{rot}}^2}{V_{\mathrm{D}}V_{\mathrm{rot}}\left(1+\frac{1}{\gamma_0}\pi_{\mathrm{c}}^{\frac{\gamma_0-1}{\gamma_0}}\right)} \qquad (10)$$

### 3.1.3 Parameters at station 1

According to the Chapman-Jouguet theory [36], the static pressure, density, relative velocity and relative Mach number of the flow at the CJ plane after detonation wave are respectively

$$p_1 = \frac{\gamma_0+\varepsilon}{(\gamma_1+1)\varepsilon}p_{\mathrm{D}}, \quad \rho_1 = \frac{\gamma_0(\gamma_1+1)}{\gamma_1(\gamma_0+\varepsilon)}\rho_{\mathrm{D}}, \quad V_1 = \frac{\gamma_1(\gamma_0+\varepsilon)}{\gamma_0(\gamma_1+1)}V_{\mathrm{D}}, \quad Ma_1 = \frac{V_1}{c_1}=1 \qquad (11)$$

where $\varepsilon = 1/Ma_{\mathrm{D}}^2$, and $V_{\mathrm{D}} = Ma_{\mathrm{D}}c_{\mathrm{D}}$ is the relative velocity of flow in front of detonation wave.

Then the $x$- and $y$-components of the absolute velocity are respectively

$$V'_{1x} = V_1\cos\theta_{\mathrm{D}} - V_{\mathrm{rot}}, \quad V'_{1y} = V_1\sin\theta_{\mathrm{D}} \qquad (12)$$

The absolute Mach number is

$$Ma'_1 = \frac{\sqrt{(V'^2_{1x}+V'^2_{1y})}}{c_1} = \sqrt{\left(\cos\theta_{\mathrm{D}} - \frac{V_{\mathrm{rot}}}{V_1}\right)^2 + \sin^2\theta_{\mathrm{D}}} \qquad (13)$$

### 3.1.4 Parameters at station 2

Assume the expansion of burned gas after detonation wave is isentropic and is complete at the outlet (that is $p_2 = p_0$), and the specific heat ratio remains unchanged during the expansion process. Then the expansion ratio $\pi_{\mathrm{e}}$ is expressed as

$$\pi_{\mathrm{e}} = \frac{p_1}{p_0} = \left(\frac{1+\frac{\gamma_1-1}{2}Ma_2^2}{1+\frac{\gamma_1-1}{2}Ma_1^2}\right)^{\frac{\gamma_1}{\gamma_1-1}} = \left(\frac{1+\frac{\gamma_1-1}{2}Ma_2^2}{1+\frac{\gamma_1-1}{2}}\right)^{\frac{\gamma_1}{\gamma_1-1}} \qquad (14)$$

The outlet flow density, temperature, relative Mach number and relative velocity are respectively

$$\rho_2 = \pi_{\mathrm{e}}^{-\frac{1}{\gamma_1}}\rho_1, \quad T_2 = T_1\pi_{\mathrm{e}}^{-\frac{\gamma_1-1}{\gamma_1}}, \quad Ma_2 = \sqrt{\frac{\gamma_1+1}{\gamma_1-1}\pi_{\mathrm{e}}^{\frac{\gamma_1-1}{\gamma_1}} - \frac{2}{\gamma_1-1}}, \quad V_2 = Ma_2\sqrt{\gamma_1\frac{p_2}{\rho_2}} \qquad (15)$$

Since the absolute velocity of the outlet flow is assumed to have only an axial ($y$-direction) component, the absolute velocity $V_{\mathrm{out}}$ and absolute Mach number $Ma_{\mathrm{out}}$ of the outlet flow are respectively

$$V_{\mathrm{out}} = \sqrt{V_2^2 - V_{\mathrm{rot}}^2}, \quad Ma_{\mathrm{out}} = Ma_2\frac{V_{\mathrm{out}}}{V_2} \qquad (16)$$

### 3.1.5 Variation of flow parameters

Define dimensionless parameters $\alpha$ and $\beta$ to represent the ratio of relative and absolute inlet Mach number to the Mach number of CJ detonation under the inlet condition, respectively, as follows

$$\alpha = \frac{Ma_0}{Ma_{D@0}}, \quad \beta = \frac{Ma_{in}}{Ma_{D@0}}, \quad \text{where } Ma_{D@0} = \frac{1}{c_0}\sqrt{2(\gamma_1^2 - 1)q} \qquad (17)$$

Assume that the inlet flow is premixed H$_2$/Air mixture with the equivalent ratio of $\phi$, and the enthalpy difference $q$ before and after combustion is approximately proportional to $\phi$, that is, $q = q^o\phi$. Table 1 shows the values of flow constants employed in the theoretical analysis. It should be noted that the specific heat ratio $\gamma_1$ after the detonation wave changes for different inlet parameters, and $\gamma_1 = 1.2$ is selected and fixed as the typical value for analysis.

**Table 1.** Flow constants employed in theoretical analysis

| Parameter | $p_0$ | $T_0$ | $\gamma_0$ | $R_0$ | $q^o$ | $\gamma_1$ | $R_1$ |
|---|---|---|---|---|---|---|---|
| Value | 100 kPa | 300 K | 1.4 | 397.8 J/(kg·K) | 5.46 MJ/kg | 1.2 | 347.9 J/(kg·K) |

Next, the variation law of flow parameters in RRDE with $\alpha$, $\beta$ and $\varphi$ are analyzed.

Some flow parameters are only functions of $\alpha$ and $\phi$, including the compression ratio $\pi_c$, expansion ratio $\pi_e$, detonation wave Mach number $Ma_D$, static pressure before and after detonation wave, temperature and other parameters at the inlet and outlet station. When $\alpha > 1$, the inlet flow is compressed and stagnate to $Ma_D$, and its static temperature increases to $T_D$. Since $Ma_D$ decreases with the increase of $T_D$, the compression ratio $\pi_c$ changes evidently with $\alpha$. As shown in Figure 4, when $\phi = 1$, and $\alpha$ increases from 1 by 1.15, $\pi_c$ rapidly increases from 1 to 40, and $Ma_D$ decreases from 5.36 to 3.17. At the same time, the static pressure $p_1$, static temperature $T_1$ and the expansion ratio $\pi_e$ after the detonation wave increase, and the static temperature $T_2$ at the outlet decreases. With the decrease of $\phi$, $Ma_D$ decreases, and the parameters such as $\pi_c$, $\pi_e$, $p_1$, $T_1$, and $T_2$ decrease correspondingly.

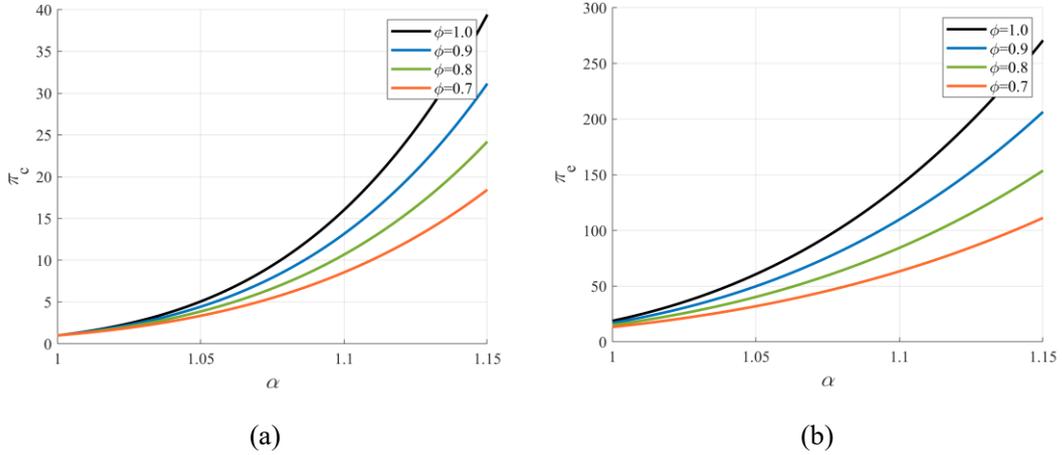

(a)      (b)

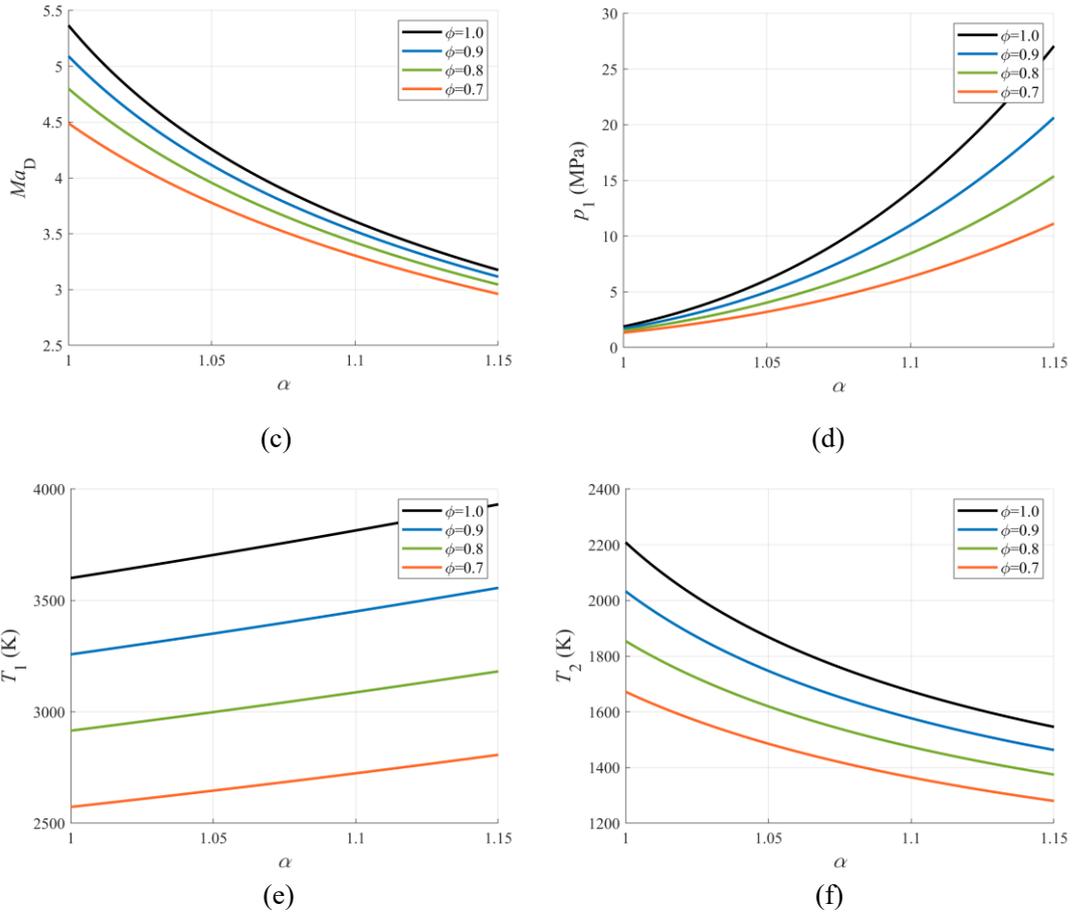

**Figure 4**. Variation of flow parameters with $\alpha$ and $\phi$. (a) $\pi_c$; (b) $\pi_e$; (c) $Ma_D$; (d) $p_1$; (e) $T_1$; (f) $T_2$.

As shown in Figure 5, if $\phi$ is fixed at 1, $V_{rot}$ and $\theta_D$ are functions of $\alpha$ and $\beta$. $V_{rot}$ monotonically increases with the increase of $\alpha$ and the decrease of $\beta$, while $\theta_D$ monotonically increases with the increase of $\alpha$ and $\beta$. For $\alpha = 1$, $\theta_D$ increases from 0 to 90° when $\beta$ increases from 0 to 1.

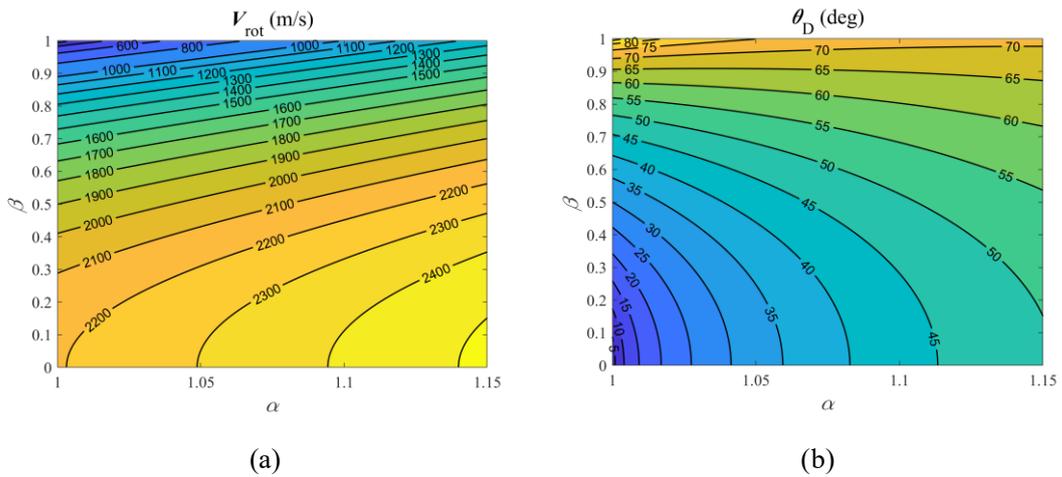

**Figure 4**. Variation of $V_{rot}$ and $\theta_D$ with $\alpha$ and $\beta$ when $\phi = 1$. (a) $V_{rot}$; (b) $\theta_D$.

### 3.2 Propulsion performance of RRDE

The flow constants in Table 1 are still used in this section to analyze the variation law of RRDE propulsion performance with inlet flow parameters ($\alpha$, $\beta$ and $\phi$), including the total pressure gain, specific thrust and thermodynamic cycle efficiency.

**3.2.1 Pressure gain characteristic and specific thrust**

With respect to the body-fixed reference frame, the total pressure at each station of RRDE are expressed as

$$p_{t0} = \left(1 + \frac{\gamma_0 - 1}{2} Ma_{in}^2\right)^{\frac{\gamma_0}{\gamma_0 - 1}} p_0 \tag{18}$$

$$p_{tD} = \left(1 + \frac{\gamma_0 - 1}{2} Ma_D'^2\right)^{\frac{\gamma_0}{\gamma_0 - 1}} p_D \tag{19}$$

$$p_{t1} = \left(1 + \frac{\gamma_1 - 1}{2} Ma_1'^2\right)^{\frac{\gamma_1}{\gamma_1 - 1}} p_1 \tag{20}$$

$$p_{t2} = \left(1 + \frac{\gamma_1 - 1}{2} Ma_{out}^2\right)^{\frac{\gamma_1}{\gamma_1 - 1}} p_2 \tag{21}$$

Then the total pressure gain PG of RRDE is expressed as

$$PG = \frac{p_{t2}}{p_{t0}} - 1 \tag{22}$$

The ratios of total pressure between station 1, station D, and station 2 are

$$\Pi_{1D} = \frac{p_{t1}}{p_{tD}}, \quad \Pi_{12} = \frac{p_{t1}}{p_{t2}} \tag{23}$$

The velocity increment (or specific thrust) $\Delta V$ at the RRDE outlet is

$$\Delta V = V_{out} - V_{in} \tag{24}$$

As shown in Figure 6, with $\alpha = 1$ and $\phi = 1$ fixed, the total pressure gain PG reaches a maximum value of 3.25 when $\beta = 0$ (i.e., the inlet flow is stationary). With the increase of $\beta$, PG monotonically decreases, and PG decreases to negative when $\beta > 0.33$, indicating that the increase of the absolute inlet velocity will reduce the total pressure gain in RRDE. When $\beta = 1$, that is, the detonation wave is standing in the body-fixed reference frame, PG decreases to −0.958. Figure 6 also shows that PG decreases as $\phi$ decreases when $\beta$ is small, while it may increase as $\phi$ decreases when $\beta$ is large.

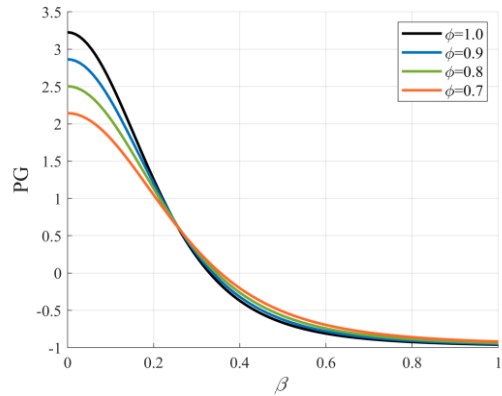

**Figure 6**. Variation of total pressure gain PG of RRDE with $\beta$ when $\phi = 1$ and $\alpha = 1$.

The variations of total pressure gain and specific thrust of RRDE with $\alpha$ and $\beta$ are further analyzed in Figure 7 with $\phi = 1$ fixed. As shown in Figure 7a, PG rapidly monotonically increases with the increase of $\alpha$ and the decrease of $\beta$, indicating that the total pressure gain of RRDE can be greatly improved by increasing the rim velocity of rotor. It should be noted that since the value of $\gamma_1$ has a great influence on the calculated results of PG, herein only the parametric influence rules of PG are discussed. The PG in real RRDE may be deviated from the calculated results shown in this section.

As shown in Figure 7b, the ratio of total pressure $\Pi_{1D}$ across the detonation wave is a monotonic function of $\beta$. While, when $\beta$ is small, $\Pi_{1D}$ first decreases and then increases with the increase of $\alpha$. Similar to the PG, when $\beta$ exceeds the critical value, $\Pi_{1D}$ is less than 1, indicating that there is no longer positive gain in the total pressure across the detonation wave. As shown in Figure 7c, the burned gas undergoes rapid expansion and performs work on the surroundings, resulting in a decrease in total pressure. A portion of the burned gas performs work to drive the propagation of the detonation wave, while the other portion of gas performs work to drive the rotation of the rotor.

The specific thrust $\Delta V$ is positively correlated with PG, and its variation with $\alpha$ and $\beta$ follows a similar pattern with PG, as shown in Figure 7d. When $\alpha = 1$, $\Delta V$ decreases from 1582 m/s to 512 m/s as $\beta$ increases from 0 to 1. When $\beta = 0$, $\Delta V$ increases from 1582 m/s to 2289 m/s as $\alpha$ increases from 1 to 1.15.

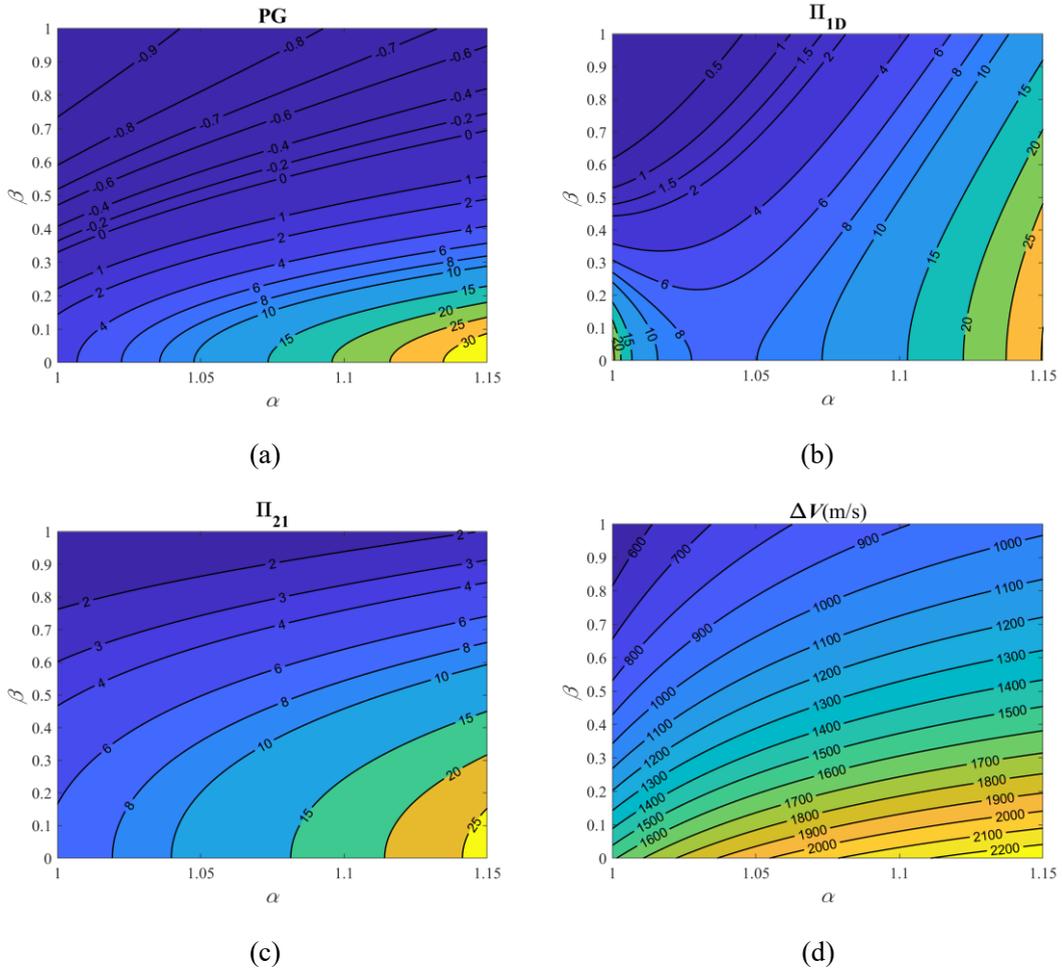

**Figure 7.** Variation of pressure gain characteristic and specific thrust of RRDE with $\alpha$ and $\beta$ when $\phi = 1$. (a) PG; (b) $\Pi_{1D}$; (c) $\Pi_{21}$; (d) $\Delta V$.

### 3.2.2 Thermodynamic cycle efficiency

Figure 8a illustrates the thermodynamic cycle in the RRDE, where 0→D represents an isentropic compression process, D→1' represents a shock compression process, 1'→1 represents the Rayleigh heat addition process, 1→2 represents the isentropic expansion process, and 2→0 represents the isobaric heat rejection process.

The efficiency $\eta_{th}$ of this thermodynamic cycle is calculated as follows,

$$\eta_{th} = 1 - \frac{Q_{2\to 0}}{Q_{D\to 1}} = 1 - \frac{\int_{T_0}^{T_2} C_P dT}{q} = 1 - \frac{\bar{C}_P (T_2 - T_0)}{q} \tag{25}$$

where $Q_{D\to 1}$ is the heat release in detonation combustion process, and $Q_{2\to 0}$ is the heat release in isobaric heat rejection process. Due to the significant decrease in gas temperature during the heat rejection process, the specific heat ratio $\gamma$ and the specific heat at constant pressure $C_P$ change significantly. For ease of analysis, the average specific heat at constant pressure $\bar{C}_P$ is selected herein to calculate the thermal efficiency, and its value is determined as follows,

$$\bar{C}_P = \frac{\bar{\gamma}}{\bar{\gamma}-1} R_1, \quad \text{where } \bar{\gamma} = \frac{\gamma_0 + \gamma_1}{2} \tag{26}$$

From Equation (25), it can be seen that when the inlet parameters are fixed, $\eta_{th}$ is mainly influenced by the outlet static temperature $T_2$. Therefore, $\eta_{th}$ is only a function of $\alpha$ and $\phi$, and does not vary with $\beta$. In the ramjet engines using isobaric combustion, the thermal efficiency can be improved by stagnating inlet flow to increase the static pressure before combustion. However, in the RRDE, in order to ensure that the detonation wave is stationary relative to the blades, the kinetic energy of this portion of the flow cannot be utilized to improve the thermal efficiency. Therefore, increasing $\beta$ is not beneficial for improving the thermodynamic cycle efficiency of RRDE, just similar to the effect of $\beta$ on PG. As shown in Figure 8b, $\eta_{th}$ monotonically increases with increasing $\alpha$ and $\phi$. When $\phi = 1$, $\eta_{th}$ increases from 0.475 to 0.655 as $\alpha$ increases from 1 to 1.15.

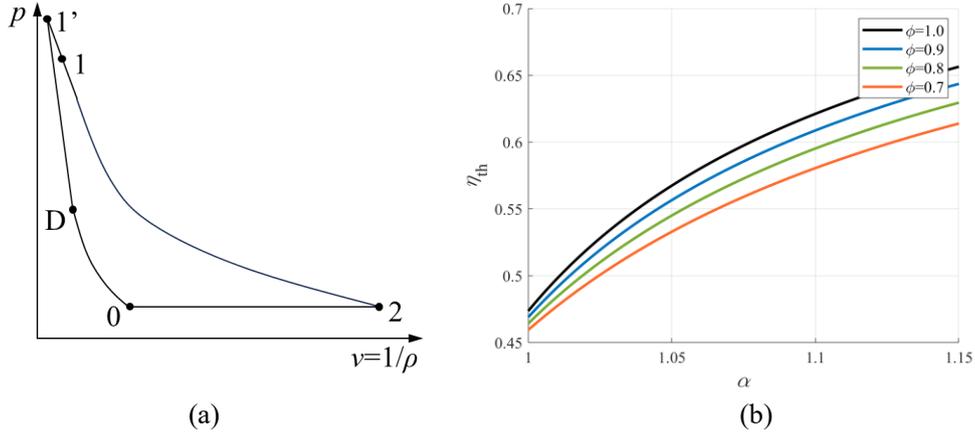

(a) (b)

**Figure 8**. (a) Schematic diagram of the RRDE thermodynamic cycle and (b) variation of thermodynamic cycle efficiency with $\alpha$ and $\phi$.

## 4. Numerical verification

The feasibility of RRDE is further examined through numerical simulations in this section, and a preliminary analysis of its performance is performed.

### 4.1 Physical model

The mid-height cross-section of the flow channel of RRDE is selected and unfolded along the circumferential direction to obtain the two-dimensional computation domain as shown in Figure 9. A moving reference system with translational velocity of $-V_{rot}$ is selected to keep the blade relatively stationary. The domain where $y \leq 0$ is the external intake stage, with periodic boundaries on the left and right sides, and a supersonic inlet boundary at the entrance boundary ($y = -H_0$). The inlet flow is the fully premixed mixture of H$_2$/Air with an equivalence ratio of $\phi$. The x-component of the inlet velocity $V_0$ is set as $V_{rot}$, the y-component is $V_{in}$, the static pressure is $p_0$, and the static temperature is $T_0$. The domain where $y > 0$ is the internal flow channel of RRDE, with wall boundaries on the left and right sides, and a supersonic outlet boundary at the exit boundary ($y = H$). The inviscid flow is adopted in the simulations. The geometric and flow parameters used in the simulations are shown in Table 2 and Table 3.

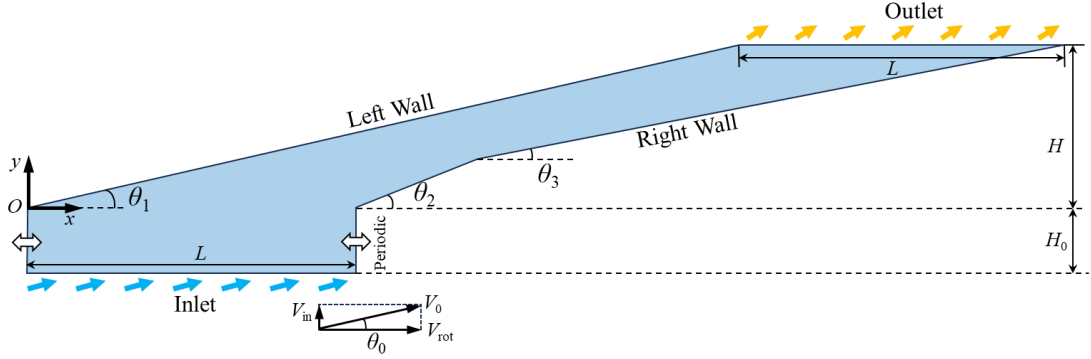

**Figure 9**. Two-dimensional computation domain used in numerical simulations.

**Table 2**. Geometric parameters of the computation domain

| Parameter | L | H | $H_0$ | $\theta_1$ | $\theta_2$ | $\theta_3$ |
|---|---|---|---|---|---|---|
| Value | 100 mm | 50 mm | 20 mm | 12.3 deg | 21.9 deg | 10.9 deg |

**Table 3**. Inlet flow parameters employed in simulations

| Parameter | $V_{rot}$ | $V_{in}$ | $p_0$ | $T_0$ | $\theta_0$ | $\phi$ |
|---|---|---|---|---|---|---|
| Value | 1549 m/s | 341 m/s | 113 kPa | 310 K | 12.3 deg | 0.318 |
|  |  |  |  |  |  | 0.353 |
|  |  |  |  |  |  | 0.388 |

## 4.2 Numerical method

In the study, a finite volume based compressible Euler equation is utilized to perform the two-dimensional numerical simulations. The compressible Euler equations are as follows

$$\frac{\partial \rho}{\partial t} + \nabla \cdot (\rho \mathbf{u}) = 0 \tag{27}$$

$$\frac{\partial}{\partial t}(\rho \mathbf{u}) + \nabla \cdot (\rho \mathbf{u}\mathbf{u}) + \nabla p = 0 \tag{28}$$

$$\frac{\partial}{\partial t}(\rho E) + \nabla \cdot (\rho E \mathbf{u} + p \mathbf{u}) = \nabla \cdot \mathbf{q} \tag{29}$$

$$\frac{\partial}{\partial t}(\rho Y_k) + \nabla \cdot (\rho Y_k \mathbf{u}) = \nabla \cdot (\rho D_k \nabla Y_k) + \dot{\omega}_k \tag{30}$$

where $\rho$, $p$ and $T$ are the density, pressure, and temperature, respectively. $\mathbf{u}$ is the velocity. $Y_k$ is the mass fraction of species $k$. $E$ is the total energy which is the sum of internal energy and kinetic energy of all components. $\mathbf{q}$ is the thermal diffusion flux consisting of thermal diffusion caused by temperature gradient and component diffusion. $\dot{\omega}_k$ is the combustion source term calculated by the finite reaction rate model. The employed reaction mechanism and the optimized Arrhenius parameters are provided in Table 4.

Table 4. Reaction and optimized Arrhenius parameters (units: m, s, mol and J/mol)

| Reaction | $A_0$ | $n$ | $E_a$ | Reaction Order |
|---|---|---|---|---|
| $H_2+0.5O_2 \rightarrow H_2O$ | $9.87 \times 10^8$ | 0 | $4.65 \times 10^7$ | $[H_2]^1 [O_2]^1$ |

The convection terms of Equations (27)-(30) are solved using the Roe-based Riemann solver and discretized using the third-order MUSCL scheme [37]. The adaptive mesh refinement (AMR) technique is applied to capture the detailed structures of detonation front. The mesh is adaptively refined based on the normalized density gradient during each time iteration [38]. The above methods and numerical code have been validated and successfully used for our previous studies on the compressible reacting flow [39,40].

The base grid size used in simulations is uniformly set to 0.3 mm. To verify the grid independency, the refinement levels of grid are respectively tested with 4, 5, and 6 levels, allowing for local grid size refinement ranging from 0.3 mm to 18.75, 9.38, and 4.69 $\mu$m in the vicinity of shock and detonation waves.

The temperature and pressure distributions on the left wall of the computation domain with different refinement levels are shown in Figure 10. Obviously, significant changes in the temperature and pressure peak locations on the left wall are observed when the refinement levels increased from 4 to 5, indicating a change in the flow field structure. However, when the refinement levels increased from 5 to 6, the changes in the flow field become very small, indicating that a 5-level refinement grid is sufficient to ensure the simulation accuracy. Therefore, the base grid size of 0.3 mm and the refinement level of 5 are adopted in the following simulations.

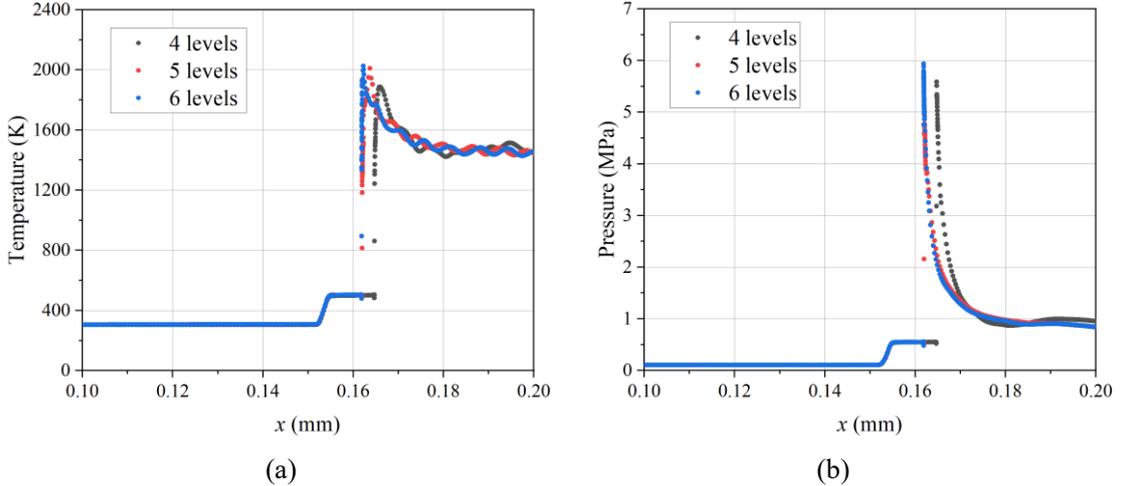

**Figure 10.** (a) Temperature and (b) pressure distributions on the left wall of the computation domain with different refinement levels.

### 4.3 Simulation results and discussions
#### 4.3.1 Flow field structure and propulsion performance

Firstly, the flow field structure and propulsion performance of the RRDE flow field are analyzed with $\phi$ = 0.353 fixed.

After the successful initiation of the detonation wave, the flow field is stabilized for 3 ms to ensure the stable standing of detonation wave. The Mach number, pressure, temperature, and total pressure contour of the flow field are shown in Figure 11. The Mach number is calculated in the moving reference system, while the total pressure is calculated in the body-fixed reference system by subtracting $V_{rot}$ from the velocity. From Figure 11a, it can be seen that the inlet Mach number is approximately 4.2. Since the inlet flow is parallel to the left wall, the flow is not compressed near the left wall. An oblique shock wave (OSW1) is induced at the right wall. The flow firstly decelerates to Ma3.6 after OSW1, and then accelerates again to around Ma4.2 after the Prandtl-Meyer flow at the corner of the right wall, thus forming a standing primary detonation wave (DW1) at $x$ = 0.161 m. OSW1 reflects at the downstream left wall to form a reflected oblique shock wave (OSW2). After OSW2, the Mach number of the flow decreases to around Ma3, and the static pressure and temperature increase significantly, and thus a secondary detonation wave (DW2) is formed downstream. The static temperature of the combustion gas rises to approximately 2100K after DW1 and DW2, and then the gas expands, leading to a rapid decrease in static pressure and temperature (Figure 11b and c).

As shown in Figure 11d, the total pressure of the flow is approximately 0.2 MPa at the inlet. The total pressure slightly increases after compression by OSW1, and then increases significantly after the detonation wave and reaches 6~8 MPa, which is 30~40 times the inlet total pressure. Subsequently, as the burned gas expands and performs work on the detonation wave and blades in the body-fixed reference frame, the total pressure of the gas rapidly decreases.

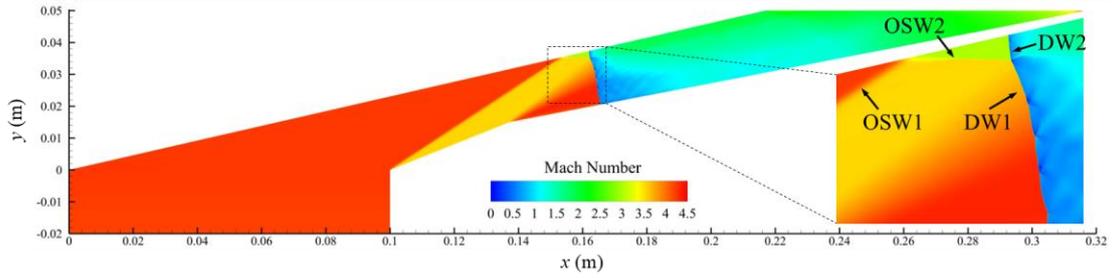

(a)

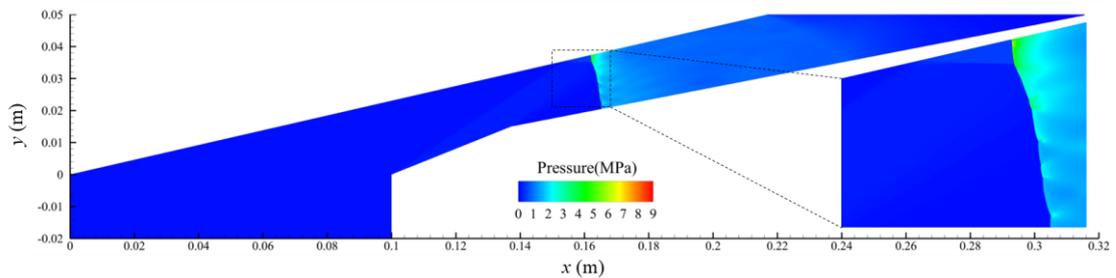

(b)

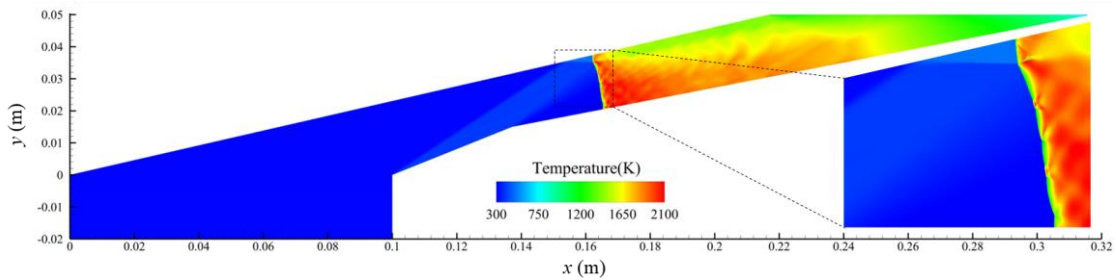

(c)

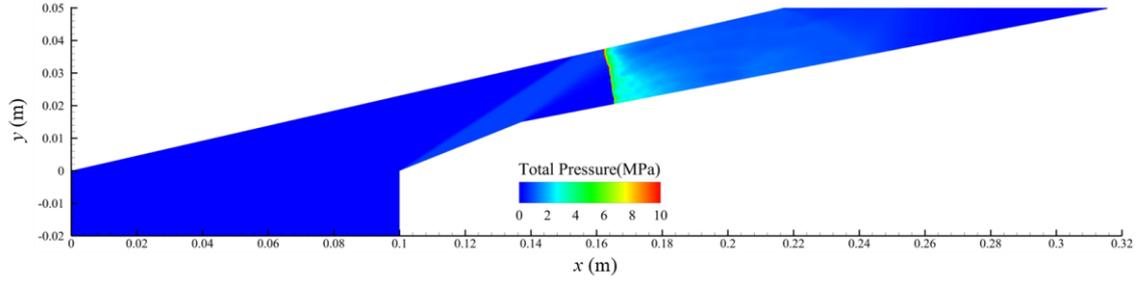

(d)

**Figure 11**. Contours of (a) Mach number, (b) static pressure, (c) temperature and (d) total pressure in RRDE with $\phi = 0.353$ fixed.

Figure 12 shows the distribution of static pressure and total pressure at the inlet and outlet of the RRDE. The left boundaries of the inlet and outlet are aligned. Because the profile design of the expansion stage adopted in the study is not optimized, the gas does not fully expand at the outlet, as shown in Figure 12a. In the region of outlet boundary where $x > 0.06$m, the gas undergoes over-expansion, resulting in a non-isentropic expansion process. The distribution of total pressure at the outlet is similar to that of static pressure. The mass-weighted average total pressure at the outlet is 0.50 MPa, which is approximately 2.6 times the inlet total pressure, and the corresponding total pressure gain PG is 1.6.

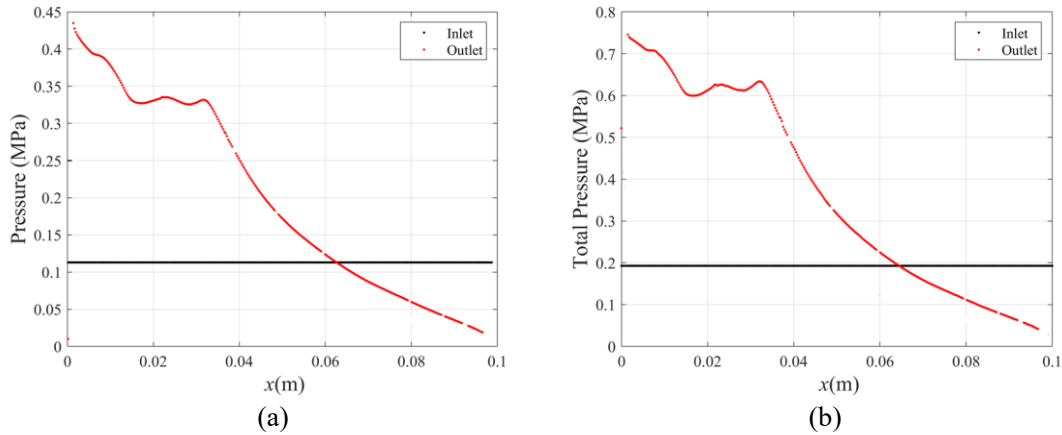

(a)            (b)

**Figure 12**. Distribution of (a) static pressure and (b) total pressure at the inlet and outlet of RRDE. The left boundaries of the inlet and outlet are aligned.

To further analyze the propulsion performance of RRDE, the distribution of wall static pressure along the $x$ and $y$ directions are plotted in Figure 13. The pressure on the right wall slightly increases after OSW1 and then rises to approximately 15 times the inlet pressure after DW1 before rapidly decreasing. The pressure on the left wall increases to 5 times the inlet pressure after OSW2, and then rises to 15 times the inlet pressure after DW2. Integrating the pressures on the left and right walls, the $x$-component (circumferential torque) of the resultant force acting on a blade with a unit height (1 m) is 3.3 kN, while the $y$-component (axial thrust, and the minus sign means positive thrust) is -21.9 kN. If the gas can fully expand at the outlet, the axial thrust can be further increased.

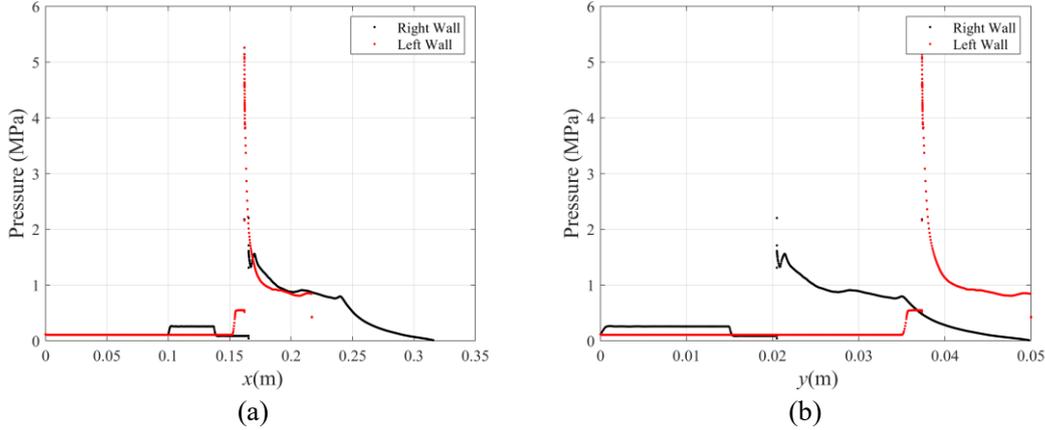

**Figure 13**. Distribution of wall static pressure along the (a) *x* and (b) *y* directions.

### 4.3.2 Influence of equivalence ratio

In order to further validate the adaptability of RRDE to the variations in the inlet parameters, the inlet equivalence ratio $\phi$ is changed to study its impact on the flow field. The results indicate that RRDE exhibits a certain resistance to disturbances in the inlet parameters. When $\phi$ varies between 0.318 and 0.388, the detonation wave can maintain a stable standing state, although there are some changes in the flow field structure.

As shown in Figure 14, when $\phi$ decreases to 0.318, the standing position of the detonation wave moves downstream. The height of DW1 decreases, while the height of DW2 increases, leading to the formation of a shock wave (SW) that is loosely coupled with the reaction zone. The shock wave here is induced by the lateral expansion of DW1. As $\phi$ continues to decrease, the standing detonation wave continues to move downstream. When the height of DW1 decrease to zero, the above flow field structure cannot be maintained, and the detonation wave eventually extinguishes.

When $\phi$ increases, the standing position of the detonation wave moves upstream, with the increase of the height of DW1 and decrease of the height of DW2. The flow field structure at $\phi = 0.388$, as shown in Figure 15, is similar to that at $\phi = 0.353$. Due to the presence of the Prandtl-Meyer flow region after OSW1, the distribution of the Mach number is non-uniform, resulting in a curved front of DW1. As $\phi$ continues to increase and DW1 crosses the reflection point of OSW1 on the left wall, the detonation wave will continuously propagate forward to the inlet because the local Mach number is lower than the CJ detonation Mach number, and the RRDE will destabilize.

It should be noted that the blade profile used in this study is only a preliminary design, and the RRDE after optimizing the blade profile should be adaptive to a wider range of inlet parameters.

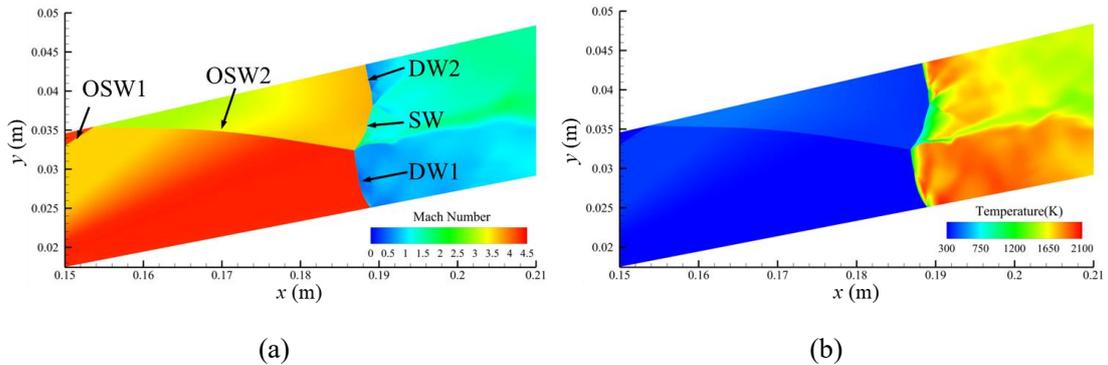

**Figure 14**. Contours of Mach number and temperature and total pressure in the vicinity of detonation wave with $\phi = 0.318$.

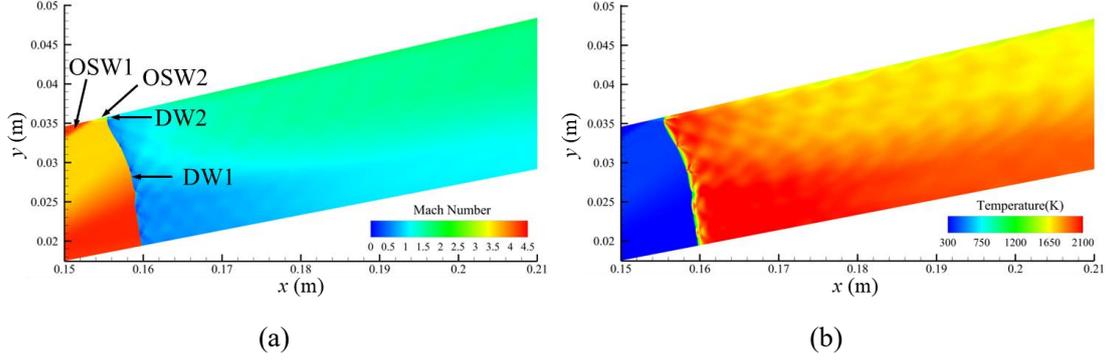

(a)                              (b)

**Figure 15.** Contours of Mach number and temperature and total pressure in the vicinity of detonation wave with $\phi = 0.388$.

The variations of propulsion performance with $\phi$ are further analyzed in Figure 16. The results show that for the RRDE used in present study, the mass-weighted total pressure gain, circumferential torque, and axial thrust do not vary monotonically with $\phi$. This is mainly because the variation of $\phi$ not only affects the heat release but also alters the detailed structure of the flow field, thereby comprehensively impacting the propulsion performance. The specific variation patterns require further detailed research.

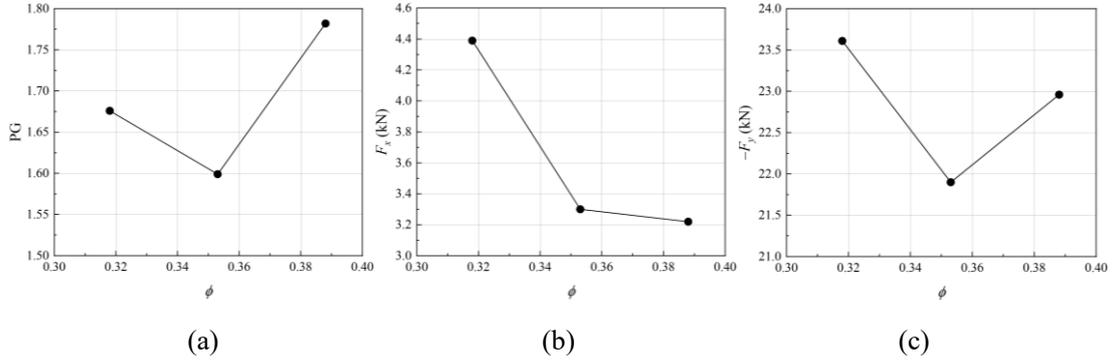

(a)                          (b)                         (c)

**Figure 16.** Variation of (a) mass-weighted total pressure gain, (b) circumferential torque, and (c) axial thrust of RRDE with the equivalence ratio $\phi$.

## 5. Conclusion

This study conducts preliminary research on a new type of detonation engine called the Ram-Rotor Detonation Engine (RRDE). The performance of RRDE and its variation laws are theoretically analyzed. The feasibility of RRDE is also verified through numerical simulations. The main conclusions obtained are as follows.

The proposed RRDE completes processes including reactant compression, combustion, and expansion within a single rotor. It utilizes the propagating detonation wave that remains stationary relative to the rotor, to enhance the total pressure gain and thermodynamic cycle efficiency. It also reduces the distance required for complete combustion, and is applicable across a wide range of inlet Mach numbers.

The theoretical analysis indicates that the performance of the RRDE is primarily influenced by the relative inlet velocity ($V_0$), the absolute inlet velocity ($V_{in}$), and the equivalence ratio ($\phi$). $V_0$ represents the combined velocity considering the blade rim velocity and $V_{in}$. The total pressure gain (PG) of RRDE and the specific thrust both decrease monotonically with increasing $V_{in}$, while they increase monotonically with increasing $V_0$. Increasing $\phi$ can improve the PG at lower $V_{in}$, but it can reduce the PG at higher $V_{in}$. The calculations show that for the premixed stoichiometric $H_2$/Air, when $V_0$ is equal to the Chapman-Jouguet detonation velocity ($V_{CJ}$) and $V_{in}$ is equal to 0, the PG can reach 3.25, and the specific thrust can reach 1582 m/s. When $V_{in} > 0.33V_{CJ}$, the PG decreases to negative and decreases rapidly with increasing $V_{in}$. The thermodynamic cycle efficiency of RRDE is primarily influenced by $V_0$ and $\phi$, and it increases monotonically with the increase of both parameters. When $\phi = 1$ and $V_0$ is increased from $V_{CJ}$ to $1.15V_{CJ}$, the efficiency increases from 0.475 to 0.655.

Numerical analysis confirms the feasibility of RRDE by demonstrating that the detonation wave can stably stand in the RRDE under appropriate inlet conditions. Additionally, the flow field exhibits the ability to adapt to changes in inlet parameters, such as the equivalence ratio. For the premixed $H_2$/Air mixture with an equivalence ratio of 0.353, under the conditions of an inlet absolute velocity of 341 m/s and a blade rim velocity of 1549 m/s, the simulated total pressure gain of RRDE can reach 1.6.

Although RRDE shows superior performance theoretically, it still faces several challenges that need to be addressed. At the mechanism level, there are concerns regarding the three-dimensional curvature effect, the initiation and stabilization mechanism of the detonation wave, stable propagation mechanism of low-velocity detonation waves, and supersonic boundary layer interference, etc. At the engineering level, there are also challenges related to high-speed rotor implementation, engine thermal protection, etc. These issues require further research in the future to ultimately achieve a high-performance RRDE.

**Acknowledgement**

The authors thank to the support from National Natural Science Foundation of China (NSFC, Grant No. 52306152) and China Postdoctoral Science Foundation (Grant No. 2023M731912).**References**

[1] S. Turns, An Introduction to Combustion: Concepts and Applications, 4th ed., McGraw-Hill, New York, 2020.

[2] Ya.B. Zeldovich, To the Question of Energy Use of Detonation Combustion, Journal of Propulsion and Power. 22 (2006) 588–592. https://doi.org/10.2514/1.22705.

[3] J.A. Nicholls, H.R. Wilkinson, R.B. Morrison, Intermittent Detonation as a Thrust-Producing Mechanism, Journal of Jet Propulsion. 27 (1957) 534–541. https://doi.org/10.2514/8.12851.

[4] R. Dunlap, R.L. Brehm, J.A. Nicholls, A Preliminary Study of the Application of Steady-State Detonative Combustion to a Reaction Engine, Journal of Jet Propulsion. 28 (1958) 451–456. https://doi.org/10.2514/8.7347.

[5] D.T. Pratt, J.W. Humphrey, D.E. Glenn, Morphology of standing oblique detonation waves, Journal of Propulsion and Power. 7 (1991) 837–845. https://doi.org/10.2514/3.23399.

[6] E. Dabora, J.-C. Broda, Standing normal detonations and oblique detonations for propulsion, in: 29th Joint Propulsion Conference and Exhibit, American Institute of Aeronautics and Astronautics, 1993. https://doi.org/10.2514/6.1993-2325.

[7] J.A. Nicholls, R.E. Cullen, K.W. Ragland, Feasibility studies of a rotating detonation wave rocket motor., Journal of Spacecraft and Rockets. 3 (1966) 893–898. https://doi.org/10.2514/3.28557.

[8] S. Eidelman, W. Grossmann, I. Lottati, Review of propulsion applications and numerical simulations of the pulsed detonation engine concept, Journal of Propulsion and Power. 7 (1991)19